\documentclass{elsart1p}
\usepackage{graphics}
\usepackage{graphicx}
\usepackage{epsfig}
\usepackage{amssymb}
\usepackage{amsmath}
\usepackage{bm}

\begin{document}

\begin{frontmatter}
\title{$K^-pp$ system with chiral SU(3) effective interaction}

\author[KEK]{Akinobu Dot\'e}
\author[TUM,YITP]{Tetsuo Hyodo}
\author[TUM]{and Wolfram Weise}

\address[KEK]{High Energy Accelerator Research Organization (IPNS/KEK), \\
1-1 Ooho, Tsukuba, Ibaraki, Japan, 305-0801}

\address[TUM]{Physik-Department, Technische Universit\"at M\"unchen, \\
D-85747 Garching, Germany}

\address[YITP]{Yukawa Institute for Theoretical Physics, Kyoto University, \\
Kyoto 606--8502, Japan}

\begin{abstract}
The $K^-pp$ system is investigated using a variational approach with 
realistic two-body interactions: the Argonne v18 $NN$ potential and an energy
dependent $\bar{K}N$ effective interaction derived from chiral SU(3) 
coupled-channel dynamics. Uncertainties in subthreshold extrapolations of the
$\bar{K}N$ interaction are considered. A weakly bound  $K^-pp$ state is 
found, with a binding energy $B = (19\pm 3)$ MeV substantially smaller than 
suggested in previous calculations. The decay width
$\Gamma(K^-pp\rightarrow \pi\Sigma N)$ is estimated to range between about 
40 and 70 MeV.

\end{abstract}
\end{frontmatter}

\section{Introduction}

In the continuing quest for possible antikaon-nuclear quasibound states, the 
$K^-pp$ system figures as an important prototype \cite{YA02}. It represents 
the simplest configuration in which the strong two-body $K^-p$ attraction 
might be amplified to form a tightly bound three-body cluster.

The FINUDA measurements with stopped $K^-$ on Li and C targets \cite{Ag05} 
seemed to suggest an interpretation in terms of strongly bound $K^-pp$ 
clusters, with a binding energy larger than 100 MeV and a width around 70 
MeV. Ever since this hypothesis was launched \cite{Ag05} and subsequently 
criticized \cite{MORT06}, there have been active developments towards 
realistic calculations of such $\bar{K}NN$ systems.  
 
Two complementary approaches have so far been used in such investigations: 
the variational method \cite{YA02,YA07} and three-body coupled-channel 
Faddeev calculations \cite{SGM07,IS07}. The $K^-pp$ system emerging from 
these computations was found to be quasibound with binding energies ranging 
from about 50 to 80 MeV. At these energies the $K^-pp\rightarrow\pi\Sigma N$ 
channel is still open. Consequently relatively large widths, between 60 and 
100 MeV, were suggested. 
 
All these calculations were based on parametrized interactions constrained by
$\bar{K}N$ scattering data close to threshold and by information about the 
$\Lambda(1405)$ as a $\bar{K}N$ quasibound state. While simple local 
potentials were employed in the variational approach, separable 
approximations for the coupled-channel interactions were used in the Faddeev 
calculations. Additional constraints from the leading chiral SU(3) 
(Tomozawa-Weinberg) interaction terms were implemented in Ref.~\cite{IS07}. 
At the same time the strong sensitivity to details of the range and energy 
dependence of the subthreshold $\bar{K}N$ interaction was pointed out in 
Ref.~\cite{DW07}. In fact, the limited predictive power in all those 
exploratory calculations can primarily be traced to ambiguities in performing
off-shell subthreshold extrapolations of the $\bar{K}N$ interactions into 
regions not yet controlled by observables. 

In the strongly coupled $\bar{K}N \leftrightarrow \pi\Sigma$ system, the 
$\Lambda(1405)$ emerges as an $I = 0$ $\bar{K}N$ quasibound state embedded in
a resonant  $\pi\Sigma$ continuum, and a detailed examination of the relevant
coupled-channel dynamics is required in order to set constraints on the input
for the $K^-pp$ studies. With this in mind, our present investigation is thus
based on a new effective $\bar{K}N$  interaction \cite{HW07} systematically 
derived from the full chiral SU(3) coupled-channel approach. An important 
issue in this context is the identification of the $\bar{K}N$ ``binding" 
energy scale associated with the $\Lambda(1405)$. The observed $\pi\Sigma$ 
mass spectrum, by its strongly asymmetric form, already indicates that there 
are subtleties involved. They relate to the two-pole analytic structure of 
the coupled-channel amplitudes: a weakly bound $\bar{K}N$ state with a pole 
located just a few MeV below threshold gets dynamically entangled with a 
broad $\pi\Sigma$ resonance. The resulting $\pi\Sigma$ mass spectrum does not
directly reflect the position of the $\bar{K}N$ quasibound state. The real 
part of the subthreshold $\bar{K}N$ amplitude ends up being located at a
center-of-mass energy $\sqrt{s} \simeq 1420$ MeV. This defines the binding 
energy scale to be used when constraining the $\bar{K}N$ amplitude, rather 
than the maximum of the $\pi\Sigma$ mass spectrum which is indeed at 
$\sqrt{s} \simeq 1405$ MeV, the position traditionally identified with the 
$\Lambda^*$ resonance. 

Consequently, the binding energy of the $\Lambda(1405)$ is not to be adjusted
at 27 MeV as it is frequently done in simple potential models. It should 
instead be fixed around 12 MeV. This significantly weaker binding translates 
from the effective $\bar{K}N$ interaction into the antikaon-nuclear few-body 
systems, as we will demonstrate.  

In the following section a Hamiltonian with realistic $\bar{K}N$ and $NN$ 
interactions as input is constructed. This Hamiltonian is used together with 
a variational ansatz for the $K^- pp$ wave function to minimize the energy 
and find the lowest quasibound state. The width of this state is estimated 
taking the expectation value of the imaginary part of the $\bar{K}N$ 
effective interaction. Further details about the $\bar{K}N$ effective 
interaction, inasmuch as they relate to the present calculation, are 
described in Section 3. Calculational details are described in Section 4. 
Results are presented and discussed in Section 5, followed by concluding 
remarks in Section 6.

\section{Framework and formalism}

\subsection{Hamiltonian}

The present calculation starts from the following non-relativistic 
Hamiltonian of the interacting $\bar{K}NN$ system:
\begin{equation}
  \hat{H} = \hat{T} + \hat{V}_{NN} + {\rm Re}\,\hat{V}_{\bar{K}N} \, 
  - \hat{T}_{CM}~~.  
\end{equation}
Here $\hat{T}$ is the total kinetic energy:
\begin{equation} 
  \hat{T}=\sum_{i=1,2} \frac{\hat{\bm{p}}^2_i}{2M_N} + 
  \frac{\hat{\bm{p}}^2_K}{2m_K}~~. 
\end{equation}
The energy of the center-of-mass motion, 
\begin{equation} 
  \hat{T}_{CM}=\frac{\left( \hat{\bm{p}}_1 + \hat{\bm{p}}_2 +
  \hat{\bm{p}}_K \right)^2}{2 \, \left( 2M_N+m_K \right)} ~~,
\end{equation}
is subtracted.

As a realistic two-nucleon interaction  $\hat{V}_{NN}$ we choose the Argonne 
v18 (Av18) potential~\cite{Av18}. This interaction reproduces $NN$ scattering
phase shifts, fits deuteron data and includes the proper short-distance $NN$ 
repulsion as an important ingredient. We are searching for a bound 
$\bar{K}NN$ state with total isospin $T=1/2$ because this configuration 
(unlike the one with $T=3/2$) makes use of the strong attraction in the 
$I = 0$ $\bar{K}N$ channel. The dominant $K^-pp$ ground state configuration 
is supposed to have the two nucleons in a spin-singlet state $(S_N = 0)$ with
isospin $T_N = 1$. The driving low-energy $\bar{K}N$ interaction does not 
change the nucleon spin, so the $NN$ pair can be in singlet-even $(^1E)$ or 
singlet-odd $(^1O)$ states (with singlet-even dominating). The important 
parts of the $NN$ interaction at work in the present context are therefore 
the central pieces of the Av18 potential, namely $v^c_{ST, NN} (r)$ as given 
in Eq. (20) of Ref.~\cite{Av18}:
\begin{equation} 
  \hat{V}_{NN} = v_{1E}(r_{12}) \, \hat{P}(^1E) + v_{1O}(r_{12}) \, 
  \hat{P}(^1O)~~, 
\end{equation} 
with projectors $\hat{P}$ onto the respective channels. For practical 
calculational purposes we use a representation of these potentials in terms 
of Gaussian forms optimally fitted to the original Av18 potentials: 
\begin{eqnarray}
  v_{1E}(r) & = & \left[3.605 \, e^{-(r/0.42 \text{fm})^2} - 0.571 
  \, e^{-(r/0.74 \text{fm})^2}
  - 0.012 \, e^{-(r/1.95  \text{fm})^2}\right]\text{GeV}~~, 
  \label{Av18_1E}\\
  v_{1O}(r) & = & \left[0.90 \, e^{-(r/0.42  \text{fm})^2} + 0.25
  \, e^{-(r/0.74  \text{fm})^2}\right]\text{GeV}~~. \label{Av18_1O}
\end{eqnarray} 

The effective $\bar{K}N$ interaction 
\begin{equation}
  \hat{V}_{\bar{K}N}  =  \hat{v}(\bar{K}N_1) +  \hat{v}(\bar{K}N_2)
\end{equation}
is given in the form of an energy dependent, complex $s$-wave potential 
derived in Ref.~\cite{HW07} from the full chiral SU(3) coupled-channel 
approach. Its components in the $\bar{K}N$ channels with isospins $I = 0,1$ 
are represented as Gaussians with a common range $a_s$:
\begin{equation}
  \hat{v}(\bar{K}N_i)  = \sum_{I=0,1}  v_{\bar{K}N}^I (\sqrt{s}) \,
  \exp\left[-(r_{\bar{K} N_i} / a_s)^2 \right]\,\hat{P}_I (\bar{K}N_i)~~, 
  \label{eq:KbarNint}
\end{equation} 
where $r_{\bar{K} N_i} = |\bm{r}_{\bar{K}} - \bm{r}_i|$ is the distance 
between the antikaon and each nucleon, and $\hat{P}_I (\bar{K}N_i)$ denotes 
the isospin projectors in the $\bar{K}N$ subsystems. The complex potential 
strength $v_{\bar{K}N}^I (\sqrt{s})$ is a function of the CM energy 
$\sqrt{s}$ in the $\bar{K}N$ subsystem. The imaginary part, Im 
$v_{\bar{K}N}^I (\sqrt{s})$, describes the open 
$\bar{K}N\rightarrow \pi\Sigma$ channels. Since the behaviour and properties 
of the $\bar{K}N$ effective interaction and its role in the $\bar{K}NN$ 
system is a key issue in the present work, we reserve a separate Section 3 
for its more detailed presentation. 

\subsection{Model wave function}

The lowest energy state of the $K^-pp$ system is found by performing a 
variational calculation,
\begin{equation}
  \delta\langle\Psi | \hat{H} -E | \Psi \rangle = 0~~,\label{var}
\end{equation}
with $|\Psi\rangle$ represented by a suitably parametrized variational wave 
function.

In the present study, the energetically most favourable $K^-pp$ configuration
is assumed to have total angular momentum and parity $J^\pi=0^-$ and total 
isospin $(T, T_3)=(1/2, 1/2)$. The parity assignment includes the intrinsic 
negative parity of the antikaon. We describe this state $|\Psi\rangle$ by the
following two-component model wave function: 
\begin{equation}
  |\Psi \rangle = {\cal N}^{-1} [ \; |\Phi_+\rangle + C \, 
  |\Phi_-\rangle \; ], \label{Mix}
\end{equation}
where ${\cal N}^{-1}$ is the normalization constant and $C$ is a mixing 
coefficient. The components $|\Phi_+\rangle$ and $|\Phi_-\rangle$ have the 
following form: 
\begin{eqnarray}
  |\Phi_+ \rangle & \equiv & 
  \Phi_+ (\bm{r}_1, \bm{r}_2, \bm{r}_K) 
  \; \left| S_N = 0 \right\rangle 
  \; \left| \, \left[ \, [NN]_{T_N=1} \, \bar{K} \, \right]_{T=1/2, T_3=1/2}
  \right\rangle~~, \label{Compont:+} \\
  |\Phi_- \rangle & \equiv & 
  \Phi_- (\bm{r}_1, \bm{r}_2, \bm{r}_K) 
  \; \left| S_N = 0 \right\rangle
  \; \left| \, \left[ \, [NN]_{T_N=0} \, \bar{K} \, \right]_{T=1/2, T_3=1/2}
  \right\rangle~~.  \label{Compont:-}
\end{eqnarray}
The first, second and third terms correspond to the spatial wave function, 
the spin state of the two nucleons and the isospin state of the total system,
respectively. In both components, the spin of the nucleon pair is assumed to 
be zero ($S_N = 0$). The state $|\Phi_+\rangle$ corresponds to the dominant 
$K^-pp$ part with inclusion of $\bar{K}^0 np$ through charge exchange. The 
mixing amplitude $\langle \Phi_+|\hat{V}_{\bar{K}N} |\Phi_- \rangle$ is 
proportional to the corresponding matrix element involving the difference 
$v_{\bar{K}N}^{I=0}-v_{\bar{K}N}^{I=1}$, with $v_{\bar{K}N}^{I=0}$ typically 
twice as strong as $v_{\bar{K}N}^{I=1}$. The admixture of the 
$|\Phi_-\rangle$ component turns out to be small, typically less than 5 \%. 
The detailed ansatz for the spatial wave functions is as follows:  
\begin{eqnarray}
  \Phi_\pm (\bm{r}_1, \bm{r}_2, \bm{r}_K) 
  & \equiv &   
  F_N(\bm{r}_1) \, F_N(\bm{r}_2) \, F_K(\bm{r}_K) \,\times
  \nonumber \\
  & &
  G(\bm{r}_1, \bm{r}_2)\;  \left[ H_1(\bm{r}_1, \bm{r}_K) \, H_2(\bm{r}_2,
  \bm{r}_K)
  \; \pm
  \; H_2(\bm{r}_1, \bm{r}_K) \,H_1(\bm{r}_2, \bm{r}_K)\right]. 
  \label{SpatialWF}
\end{eqnarray}
Here, $F_N(\bm{r}_i)$ ($i=1,2$) and $F_K(\bm{r}_K)$ are trial functions 
describing the localization of the nucleon and the kaon, respectively. Their 
forms are assumed to be single Gaussians:  
\begin{equation}
    F_N(\bm{r}_i) \equiv  \exp[-\mu \, \bm{r}^2_i]~,~~~~~ 
    F_K(\bm{r}_K) \equiv  \exp[-\gamma \, \bm{r}^2_K]~~. \label{SPmotion}
\end{equation}
We introduce the $NN$ correlation function $G(\bm{r}_1, \bm{r}_2)$ and 
$\bar{K}N $correlation functions $H_\alpha(\bm{r}_i, \bm{r}_K)$ (with 
$\alpha = 1,2$ and $i =1,2$) of the following form: 
\begin{eqnarray}
    G(\bm{r}_1, \bm{r}_2) & \equiv & 
    1 - \sum_{n=1}^{N_N} 
    f^{NN}_n \exp\left[-\lambda^{NN}_n \left(\bm{r}_1 - \bm{r}_2
    \right)^2 \right] , \label{NNcorr} \\
    H_\alpha(\bm{r}_i, \bm{r}_K) & \equiv & 
    1 + \sum_{n=1}^{N_K} 
    f^{\bar{K}N}_{\alpha,n} \exp\left[-\lambda^{\bar{K}N}_n 
    \left(\bm{r}_i - \bm{r}_K
    \right)^2 \right] . \label{KNcorr}
\end{eqnarray}
The $NN$ correlation function properly accounts for the strong short-distance
repulsion in the $NN$ interaction by keeping the two nucleons apart. The 
$\bar{K}N$ correlation functions can flexibly adjust themselves to the 
antikaon-nucleon attractive interaction so as to variationally determine the 
configuration which minimizes the energy.

Note that the spatial wave functions 
$\Phi_\pm (\bm{r}_1, \bm{r}_2, \bm{r}_K)$ of Eq. (\ref{SpatialWF}) are even 
or odd under exchange of two nucleons: 
\begin{equation}
    \Phi_\pm (\bm{r}_2, \bm{r}_1, \bm{r}_K) 
    = \, \pm \, \Phi_\pm (\bm{r}_1, \bm{r}_2, \bm{r}_K)~~ .
\end{equation}
The two nucleons in $|\Phi_+\rangle$ are thus in a singlet-even state, while
in $|\Phi_-\rangle$ they are in a singlet-odd state. 

The model wave functions (\ref{Mix}-\ref{KNcorr}) have real-valued 
variational parameters: $C$ in Eq.~(\ref{Mix}), $\mu$ and  $\gamma$ in 
Eq.~(\ref{SPmotion}), $\{ f^{NN}_n, \lambda^{NN}_n \}$ ($n=1, ... , N_N$) in 
Eq. (\ref{NNcorr}), and $\{ f^{\bar{K}N}_{\alpha,n}, \lambda^{\bar{K}N}_n \}$
($n=1, ... , N_K$) in Eq. (\ref{KNcorr}). The variational principle 
(\ref{var}) then determines the optimal parameter set which minimizes the 
expectation value of the total Hamiltonian. 

\section{Effective $\bar{K}N$ potential}

Here we discuss the $\bar{K}N$ potential used in Eq.~\eqref{eq:KbarNint}. The
strangeness $S=-1$ meson-baryon scattering and the properties of the 
$\Lambda(1405)$ resonance are well described by the chiral SU(3) 
coupled-channel approach~\cite{Kaiser:1995eg,Oset:1998it,Oller:2000fj,
Lutz:2001yb}. In Ref.~\cite{HW07}, two of the present authors have derived 
the effective $\bar{K}N$ interaction based on chiral SU(3) dynamics. First, 
the coupled-channel framework is translated into the equivalent single 
$\bar{K}N$ channel problem with a complex and energy-dependent interaction 
kernel, ${\bf V}_{\text{eff}}^I(\sqrt{s})$ defined in Ref.~\cite{HW07}, which
fully incorporates the dynamics of the eliminated channels. Starting from 
this effective interaction, a local $\bar{K}N$ potential $U^I(r,\sqrt{s})$ is
then constructed in each isospin channel, to be used in the  $\bar{K}N$ 
two-body Schr\"odinger equation
\begin{equation}
    -\frac{1}{2\mu}\frac{d^2 u(r)}{dr^2}
    +U^I(r,\sqrt{s})\,u(r) = -B\,u(r)~~ ,
    \label{eq:schroedinger}
\end{equation}
where $u(r)$ is the radial $\bar{K}N$ $s$-wave function and $B$ is the 
binding energy. We adopt a Gaussian form for the spatial distribution of the 
$\bar{K}N$ potential:
\begin{equation}
    U^I(r,\sqrt{s}) 
    = v^{I}_{\bar{K}N}(\sqrt{s})
    \, \exp\left[-(r / a_s)^2 \right] ~~, 
    \label{eq:KbarNpot1}
\end{equation}
where $r$ is the relative coordinate in the $\bar{K}N$ system and $a_s$ is 
the range parameter of the potential.

The potential strength is related to the single-channel $\bar{K}N$ effective 
interaction kernel ${\bf V}_{\text{eff}}^I(\sqrt{s})$, derived from the 
coupled-channel approach, as
\begin{equation}
    v^{I}_{\bar{K}N}(\sqrt{s})
    = -\frac{4\pi}{\pi^{3/2}\,a_s^3}
    \frac{ {\bf V}^I_{\text{eff}}(\sqrt{s})}
    {2\tilde{\omega}}
    \label{eq:uncorrected}~~ ,
\end{equation}
where $\tilde{\omega}$ is the reduced energy of the $\bar{K}N$ two-body 
system. The potential so obtained is complex and energy dependent, reflecting
the elimination of the dynamics of the $\pi\Sigma$ and other (less important)
channels. The center-of-mass energy $\sqrt{s}$ is related to the binding 
energy of the two-body system as $B = M_N+m_K -\sqrt{s}$. We choose the range
parameter $a_s$ such that the resonance structure in the $I=0$ channel below 
the $\bar{K}N$ threshold matches the result of the full chiral SU(3) dynamics
calculation. Around $\bar{K}N$ threshold, the scattering amplitudes of both 
$I=0$ and $I=1$ are well reproduced by this potential. 

However, approximating the full  $\bar{K}N$ effective interaction 
${\bf V}_{\text{eff}}^I(\sqrt{s})$ by a {\it local} potential 
$ U^I(r,\sqrt{s})$ can obviously work only in a limited energy range. It was 
found indeed that a simple extrapolation of the 
potential~(\ref{eq:uncorrected}) to the deep subthreshold region, 
$\sqrt{s}<1400$ MeV, significantly overestimates the scattering amplitude in 
comparison with the full chiral dynamics result~\cite{HW07}. Extra energy 
dependence is required in the approximate local potential to repair this 
deficiency. A correction is applied, modifying the energy dependent strength 
of the real part of the potential such that the scattering amplitude of the 
full chiral dynamics calculation is reproduced all the way down to 
$\sqrt{s}=1300$ MeV. The energy dependence of these ``corrected'' potentials
is parametrized by polynomials as
\begin{equation}
    v^I_{\bar{K}N}(\sqrt{s})
    = K_{0}^I + K_{1}^I s^{1/2} + K_{2}^I\,s + K_{3}^I\, s^{3/2} .
    \label{eq:corrected}
\end{equation}
with coefficients $K_{i}^I$ given in Ref.~\cite{HW07}. These improved local 
potentials are then used in the three-body variational calculation. 

In order to estimate systematic theoretical uncertainties, we adopt as in 
Ref.~\cite{HW07} four different versions of chiral SU(3) dynamics approaches 
to construct the equivalent local potentials,ORB~\cite{Oset:2001cn}, 
HNJH~\cite{Hyodo:2002pk}, BNW~\cite{Borasoy:2005ie}, and 
BMN~\cite{Borasoy:2006sr}, all of which reproduce the experimental data of 
the $\bar{K}N$ scattering and the properties of the $\Lambda(1405)$. The 
values of their range parameters are shown in the second row of 
Table~\ref{tbl:KbarNresult}.

Let us examine more closely the structure of the two-body $\bar{K}N$ system 
with $I=0$ which features the $\Lambda(1405)$ as a quasibound state below 
$\bar{K}N$ threshold. Given the small imaginary part of the potential 
\cite{HW07}, we can solve the Schr\"odinger equation~\eqref{eq:schroedinger} 
starting with Re $U^{I=0}$ and study the structure of this quasibound state.
The solution is found self-consistently, with the energy dependence of the
potential fully taken into account. The results for the $\bar{K}N$ binding 
energies $B$ in the $I = 0$ channel are summarized in 
Table~\ref{tbl:KbarNresult} together with the mean $\bar{K}N$ distance of the
bound state wave functions. We find $B\sim 10 - 13$ MeV and 
$\sqrt{\langle r^2\rangle} \sim 1.7 - 2.0$ fm. Based on the calculations 
reported in Ref.~\cite{HW07}, we assign an estimated additional $3-4$ MeV 
uncertainty to these values of $B$ from dispersive effects induced by the 
imaginary part of the $\bar{K}N$ potential.

Note that the small binding energy $B$ has its correspondence in the zero of 
the real part of the corresponding subthreshold (off-shell)  $\bar{K}N$ 
scattering amplitude. This zero is consistently located around 
$\sqrt{s} \simeq 1420$ MeV for all four variants of chiral SU(3) models, 
{\it not} at 1405 MeV as one would naively expect. At the same time, the 
maximum of the calculated $\pi\Sigma$ invariant mass spectrum is indeed 
located around $\sqrt{s} \simeq 1405$ MeV. In chiral SU(3) coupled-channel 
dynamics, these important features are understood as originating from the 
strong $\pi\Sigma$ interaction and  discussed in detail in Ref.~\cite{HW07}. 

The results just mentioned should be compared with those of the
phenomenological model~\cite{YA07} in which a local $K^-p$ potential, 
unconstrained by chiral SU(3), is tuned to yield $B = 27$ MeV and thus 
produces a smaller size of the quasibound state, 
$\sqrt{\langle r^2\rangle}= 1.36 $ fm. Given the substantially weaker 
$\bar{K}N$ attraction in our chiral SU(3) dynamics approach, it is then
perhaps not surprising that, in the present work, the binding energy of the 
$K^-pp$ cluster will end up not far from twice the binding energy of the 
individual $K^-p$ state, at about 20 MeV as we shall demonstrate, more than a
factor of two lower than the $K^-pp$ binding energy predicted in 
Ref. \cite{YA07}.
\begin{table}[tbp]
    \centering
    \caption{ Binding energies $B$ and mean $\bar{K}N$ distance $\sqrt{r^2}$ 
    of the quasibound $\bar{K}N$ state (the $\Lambda(1405)$) calculated using
    Eq.~(\ref{eq:schroedinger}) with equivalent local $\bar{K}N$ potentials 
    derived from four variants of the chiral SU(3) coupled-channel approach. 
    The range parameters $a_s$ of the potentials are collected in the second 
    row.}

        \begin{tabular}{c r|rrrrrrr}
	
        &&ORB &&HNJH && BNW && BMN \\
	  & & \cite{Oset:2001cn} 
	& & \cite{Hyodo:2002pk} 
	& & \cite{Borasoy:2005ie} 
	& & \cite{Borasoy:2006sr} \\
        \hline
        $a_s$ [fm]  && 0.52 && 0.47 && 0.51 && 0.41 \\
        $B$ [MeV]  &&  $11.77$ && $11.47$ && $9.97$ && $13.31$ \\
        $\sqrt{\langle r^2 \rangle}$ [fm]  
	& & $1.87$ && $1.86$ && $1.99$  && $1.72$   \\
	\hline
    \end{tabular}
    \label{tbl:KbarNresult}
\end{table}

\section{Calculational procedure}

Given the explicit energy dependence of the $\bar{K}N$ potential, a 
self-consistent solution of the variational Eq.~(\ref{var}) must be found.
This is done in the same way as described in our previous work~\cite{DW07}.
In this procedure, an auxiliary antikaon ``binding energy" $B_K$ is 
introduced and defined as follows:
\begin{equation}
    -B_K \equiv 
    \langle \Psi | \hat{H} | \Psi \rangle 
    - \langle \Psi | \hat{H}_{N} | \Psi \rangle~~.
    \label{eq:BK}
\end{equation}
Here $\hat{H}_{N}$ is the Hamiltonian of the two-nucleon subsystem: 
\begin{eqnarray}
    \hat{H}_{N} & = & \hat{T}_{N} + \hat{V}_{NN} - \hat{T}_{CM, N}~~, \\
    \hat{T}_{N} & = & \frac{\hat{\bm{p}}^2_1 
    +\hat{\bm{p}}^2_2}{2M_N}~~, ~~~~~~ 
    \hat{T}_{CM, N}=\frac{\left(  \hat{\bm{p}}_1
    + \hat{\bm{p}}_2 \right)^2}{4M_N}~~.  
\end{eqnarray}
Obviously $B_K$ is not an observable since 
$\langle \Psi | \hat{H}_{N} | \Psi \rangle$ is not an observable either, but 
it is a useful variable to control the energy $\sqrt{s}$ of the $\bar{K}N$ 
subsystem as it enters the potential $\hat{V}_{\bar{K}N}$. 

The relation between $\sqrt{s}$ and $B_K$ is not a priori fixed since 
$\sqrt{s}$ is the energy of a two-body subsystem within the three-body 
system. In general,
\begin{equation}
    \sqrt{s} = M_N + m_K - \eta\, B_K  \label{DEF}~~,
\end{equation}
where $\eta$ is a parameter describing the balance of the antikaon energy 
between the two nucleons of the $\bar{K}NN$ three-body system. One expects 
$1/2 \le\eta \le1$. The upper limit ($\eta = 1$) corresponds to the case in 
which the antikaon field collectively surrounds the two nucleons, a situation
encountered in the limit of static (infinitely heavy) nucleon sources. In the
lower limit ($\eta = 1/2$) the antikaon energy is split symmetrically 
half-and-half between the two nucleons. We will investigate both cases and 
label them ``Type I" and ``Type II", respectively:
\begin{eqnarray}
    {\rm Type \; I \; :} ~~~~~~~~\sqrt{s}& = &M_N + m_K - B_K~~, 
    \label{DEF1}\\
    {\rm Type \; II \; :} ~~~~~~~~\sqrt{s}& = &M_N + m_K - B_K/2~~. 
    \label{DEF2}
\end{eqnarray}
The actual calculation now proceeds as follows. First, assume a trial 
starting value $B_K^{(0)}$ and determine $\sqrt{s}$ with either the Type I or
the Type II option. This specifies $\sqrt{s}$ in the input $\bar{K}N$ 
potential. Then perform the variational calculation to determine the minimum 
energy of the system. With the resulting wave function, calculate the 
improved antikaon binding energy $B_K^{(1)}$ according to Eq.~(\ref{eq:BK}). 
Examine whether $B_K^{(1)}$ coincides with $B_K^{(0)}$. If not, iterate this 
procedure until $B_K^{(n)} \simeq B_K^{(n-1)}$ is satisfied at a prescribed 
level of accuracy\footnote{In practice, the input $B_K^{(0)}$ is optimized by
hand as outlined in Ref.~\cite{DW07}.}.

The $K^-pp$ bound state $|\Psi\rangle$ is calculated variationally using the
real part Re$\hat{V}_{\bar{K}N}$ of the $\bar{K}N$ potential. The decay width
$\Gamma$ for $K^-pp\rightarrow\pi\Sigma N$ is then estimated in leading order
perturbation theory as 
\begin{equation} 
    \Gamma(K^-pp\rightarrow\pi\Sigma N) 
    = -2 \; \langle \Psi | \, {\rm Im} \hat{V}_{\bar{K}N} | \Psi \rangle. 
\end{equation}
Such an estimate is justified by the fact that 
Im$\hat{V}_{\bar{K}N} \ll |$Re$\hat{V}_{\bar{K}N}|$ \cite{HW07}. However, the
detailed balance between kinetic and potential energy terms finally produces 
a weakly bound, short-lived state whose binding energy is smaller than the 
width, so that this estimate of $\Gamma$ should only be taken for qualitative
orientation.  

\section{Results \label{RESULT}}

We now present results of our variational $K^-pp$ calculation. All four 
variants of $\bar{K}N$ potentials derived from chiral SU(3) dynamics 
(``ORB'', ``HNJH'', ``BNW'' and ``BMN'', as explained in Section 3) have been
used in order to estimate theoretical uncertainties. Both versions for the 
splitting relation between $\sqrt{s}$ and $B_K$, Type I (Eq. (\ref{DEF1})) 
and Type II (Eq. (\ref{DEF2})), have been employed in comparison. The results
of the self-consistent calculations are collected in Table 
\ref{tab:Res/Summary}. They all predict  weak $K^-pp$ binding, considerably 
weaker than what was found in previous computations. The total $K^-pp$ 
binding energy (BE) ranges from 15.6 MeV to 17.7 MeV for the Type I scenario 
and from 19.6 to 21.6 MeV for Type II. The detailed energy dependence of the 
$\bar{K}N$ effective interactions obviously matters, and there is an 
indication that the Type II configuration may be energetically favoured over 
Type I. The decay width for $K^-pp\rightarrow \pi\Sigma N$ induced by the 
imaginary part of $\hat{V}_{\bar{K}N}$ is in intervals $39-53$ MeV for Type I
and  $54-72$ MeV for Type II. Altogether we combine both the Type I and Type 
II scenarios in a conservative estimate of uncertainties, resulting in an 
predicted binding energy range $B = 19\pm3$ MeV and a width ranging between 
40 and 70 MeV. 

The calculated average distance between the two nucleons in the $K^-pp$ bound
state is $R_{NN} \simeq 2.2$ fm. This is obviously not a very dense system. 
The average between the antikaon and a given nucleon is $R_{\bar{K}N} \simeq 
1.9$ fm, not far from the mean $\bar{K}N$ distance of the isolated 
$\Lambda(1405)$ quasibound state (see Table \ref{tbl:KbarNresult}). 

It is instructive to examine the detailed decomposition of the total $K^-pp$ 
energy into kinetic and potential energy pieces of the $NN$ and  $\bar{K}N$ 
subsystems. This is shown in Table \ref{tab:Res/Smooth/Wf} for the 
energetically favoured Type II case. One notes that the nucleons are the 
``slow" movers in this system, with small kinetic energies per nucleon around
20 MeV in all models considered. The $\bar{K}$-nuclear potential energy is 
large and negative, but it wins over the total kinetic energy of the system 
by only a few MeV. The additional binding is then provided by the moderate 
average $NN$ potential energy  of about $-15$ MeV.
\begin{table}
\centering
\caption{
Results of the self-consistent variational $K^-pp$ calculations using 
effective $\bar{K}N$ interactions based on chiral SU(3) coupled-channel 
dynamics as explained in the text. Upper row: range $a_s$ of the effective 
potential (\ref{eq:KbarNpot1}). Shown are the total $K^-pp$ binding energy 
(BE) and the $K^-pp\rightarrow \pi\Sigma N$ decay width ($\Gamma$) for Type I
and Type II configurations (\ref{DEF1},\ref{DEF2}).}
\label{tab:Res/Summary} 

\begin{tabular}{l |cccccccc}

        &&ORB&&HNJH &&BNW &&BMN \\
$a_s$ [fm]  && 0.52& & 0.47 && 0.51   & & 0.41    \\
\hline
Type I & &  &&&&   &    &    \\
BE [MeV] & &17.7& & 15.9 & &17.1 && 15.6 \\
$\Gamma$  [MeV] & &53.2 && 47.1 && 60.9 && 39.2 \\
\hline
Type II  &   &   &    &    \\
BE [MeV] && 21.6 && 19.8 && 19.6 && 20.8 \\
$\Gamma$  [MeV] && 64.5 && 58.6 && 71.7 && 53.7 \\
\hline
\end{tabular}
\end{table}

\begin{table*}
\caption{
Detailed decomposition of the total $K^-pp$ binding energy (BE) for the Type
II configuration: binding energy ($B_K$) of the $\bar{K}N$ subsystem (see 
Eq.~(\ref{eq:BK})), energy $E_N = \langle\Psi|\hat{H}_N|\Psi\rangle$ and 
kinetic energy $T_N = \langle\Psi|\hat{T}_N - \hat{T}_{CM,N}|\Psi\rangle$ of 
the two-nucleon subsystem, total kinetic energy 
$T_{tot} = \langle\Psi|\hat{T} - \hat{T}_{CM}|\Psi\rangle$, and potential 
energies  $V(NN) = \langle\Psi|\hat{V}_{NN}|\Psi\rangle$, 
$V(KN) = \langle\Psi|\text{Re}\hat{V}_{\bar{K}N}|\Psi\rangle$.}
\label{tab:Res/Smooth/Wf} 
\centering
\begin{tabular}{l  c|ccc | cccccccc | cccccc}
& & &BE [MeV] &  & $B_K$[MeV]  & &$E_N$[MeV]  & &$T_N$[MeV] 
& &$T_{tot}$[MeV]  & &V(NN)[MeV] & &V(KN)[MeV]  \\
\hline 
ORB && & 21.6 & & 46.1 & &24.5 & &40.1 & &136.0 && $-$15.6 && $-$142.0\\
HNJH && &19.8 & & 45.2 & &25.4 && 40.3 &&141.5 && $-$14.9 & &$-$146.4 \\
BNW && & 19.6 && 43.2& & 23.6 && 38.5 && 132.1 && $-$14.9& & $-$136.8 \\
BMN && & 20.8 & & 49.6 && 28.8 && 43.4 && 160.0 && $-$14.6& & 
$-$166.1 \\
\hline
\end{tabular}

\end{table*}

\section{Summary and concluding remarks}
The present variational calculation of the $ppK^-$ system, as a prototype for
antikaon-nuclear quasibound states, has been performed with the aim to 
satisfy two important minimal requirements, namely the use of \\
\begin{itemize} 
    \item{a realistic nucleon-nucleon interaction (here: the Av 18 $NN$ 
    potential);}\\
    \item{ a realistic  $\bar{K}N$ interaction (here: the subthreshold 
    effective $\bar{K}N$ interaction based on chiral coupled-channel 
    dynamics).}\\
\end{itemize}
The variational $\bar{K}NN$ wave function has been constructed so as to 
handle the strong short-distance $NN$ interaction which keeps the two 
nucleons apart. The effective $\bar{K}N$ interaction incorporates essential 
features of the $\bar{K}N \leftrightarrow\pi\Sigma$ coupled-channel dynamics.
In particular, it accounts for the important fact that the zero in the real 
part of the $I=0$ subthreshold $\bar{K}N$ amplitude and the maximum of the 
$\pi\Sigma$ invariant mass spectrum do not coincide: the quasibound 
$\bar{K}N$ state is located around $\sqrt{s} \simeq$ 1420 MeV whereas the 
$\pi\Sigma$ mass spectrum peaks at $\sqrt{s} \simeq$ 1405 MeV. This implies 
weaker  $\bar{K}N$ attraction than naively anticipated. As a consequence, the
predicted $K^-pp$ binding energy found in the present calculations is
\begin{equation}
    B(K^-pp) \simeq (19 \pm 3)\,\text{MeV}~~,
\end{equation}
where the uncertainty measure is based entirely on using four different 
versions of chiral SU(3) coupled-channel models as input. Additional 
systematic uncertainties, such as the dispersive shift induced by the 
imaginary part of the $\bar{K}N$ potential, are under analysis~\cite{DHW08}.

The $K^-pp\rightarrow\pi\Sigma N$ decay width is estimated to be, roughly,
\begin{equation}
    \Gamma(K^-pp\rightarrow\pi\Sigma N) \sim (40 - 70)\,\text{MeV}~~.
\end{equation}
This suggests that $K^-pp$ clusters, even if quasibound, would be difficult 
to identify experimentally. The width is in fact expected to increase even 
more through the non-mesonic decay $K^-pp\rightarrow YN$ into a 
hyperon-nucleon pair \cite{WH08}. Detailed studies of this and further 
corrections (such as the influence of spin-dependent $NN$ correlations and 
the role of $p$-wave $\bar{K}N$ interactions involving the $\Sigma^*(1385)$)
are in progress and will be reported elsewhere \cite{DHW08}. 

The $K^-pp$ binding energy found in the present calculation is significantly 
smaller than corresponding values reported from variational \cite{YA02,YA07} 
and Faddeev \cite{SGM07,IS07} calculations. While the difference with 
respect to the previous variational results is understood in terms of the 
improved chiral $\bar{K}N$ interaction used in the present approach, a direct
comparison with the Faddeev results (which explicitly incorporate 
coupled-channel dynamics, though with separable potentials) is not so obvious
and requires further detailed studies.

In any case it is found that the $K^-pp$ binding energy turns out to be very 
sensitive to details of the off-shell, subthreshold extrapolation of the 
$\bar{K}N$ interaction. This extrapolation relies so far on constraints from 
threshold scattering and kaonic hydrogen measurements, together with the 
available low-statistics data of the $\pi\Sigma$ invariant mass spectrum.
Stronger constraints are expected to be imposed once kaonic hydrogen and 
deuterium precision measurements become available. A better determination of 
the $\pi\Sigma$ mass spectrum would also be highly welcome, as well as 
exclusive data on the final states from decaying antikaon-nuclear systems in 
order to clarify their dynamics.  
 
\vspace{0.8cm}
\noindent
{\bf Acknowledgements}\\
\\
\noindent
We thank Avraham Gal for fruitful and stimulating discussions during his 
visit in Munich. One of the authors (A. D.) is grateful to Prof. Akaishi for 
his advice on constructing our model wave function. This project is partially
supported by BMBF, GSI and by the DFG excellence cluster ``Origin and 
Structure of the Universe".  T.~H. thanks the Japan Society for the Promotion
of Science (JSPS) for financial support. This work is supported in part by 
the Grant for Scientific Research (No.\ 19853500, 19740163) from the Ministry
of Education, Culture, Sports, Science and Technology (MEXT) of Japan. This 
research is  part of Yukawa International Program for Quark-Hadron Sciences.

\end{document}